\documentclass[12pt,a4paper]{article} 
\usepackage[dvipsnames,usenames]{color}
\usepackage{multirow}
\usepackage{ftnxtra}
\usepackage{subeqn}
\usepackage[numbers,square,comma,sort&compress]{natbib}
\usepackage{lscape}
\usepackage{graphicx}
\usepackage{booktabs}

\begin{document}

\title{The general static spherical perfect fluid solution in GR with EoS parameter $w=-1/6$}

\author{\.{I}brahim 
Semiz\thanks{mail: ibrahim.semiz@bogazici.edu.tr, ibrahim.semiz.fizik@icloud.com} \\ \\
Bo\u{g}azi\c{c}i University, Department of Physics\\
34342 Bebek, \.{I}stanbul, TURKEY}
    
\date{ }

\maketitle

\begin{abstract}
The general analytical solution in GR for the static spherically symmetric metric supported by a perfect fluid with proportional-equation-of-state $p = w \rho$ is not known at the time of this writing, except for the trivial cases $w=0$ and $w=-1$; for $w=-1/3$, and the recently reported $w=-1/5$. We show that the case $w=-1/6$ is also analytically solvable, as predicted in another recent work. The solution is affected by a Buchdahl transformation of a known solution. We discuss the spacetimes different ranges of the solution's parameters represent.
\end{abstract}


\section{Introduction: SSSPF solutions for isothermal EoS}\label{Intro}

Despite the impression in a number of works  that cite~\cite{Kiselev}, we do not yet know the spacetime metric for a ``blackhole surrounded by quintessence''~\cite{Visser_on_Kiselev,Lake_on_Kiselev,Semiz_on_Kiselev}. That is to say, the {\it general} exact solution of Einstein's Field Equations (EFE)
\begin{equation}
G_{\mu\nu} = \kappa T_{\mu\nu}               \label{ee}
\end{equation}
are not known for
\begin{equation}
T_{\mu\nu} = (\rho + p) \, u_{\mu} u_{\nu} + p \, g_{\mu\nu}  \label{pfemt}
\end{equation}
where
\begin{equation}
p=w\rho.     \label{eos}
\end{equation}
for constant $w$, for the static spherically symmetric (SSS) case. 

In eq.(\ref{ee}), $G_{\mu\nu}$ is the Einstein tensor, and for its definition we use the conventions of~\cite{mtw}. $\kappa$ is the coupling constant, and $T_{\mu\nu}$ the stress-energy-momentum (SEM) tensor. The form (\ref{pfemt}) for $T_{\mu\nu}$ describes a so-called {\it perfect fluid}, sometimes called the {\it isotropic} perfect fluid, which corresponds to a fluid without viscosity and heat conduction, where $\rho$ and $p$ are the energy density and pressure, respectively, as measured by an observer moving with the fluid; and $u_{\mu}$ is the fluid's four-velocity. Another aspect  of the description of a perfect fluid is an assumed relation $f(p,\rho)=0$, called an {\em equation of state} (EoS). For analyses of stellar structure, the {\it polytropic} EoS, $p\propto \rho^{\gamma}$ is often used, while in cosmology, the isothermal\footnote{In the literature, this EoS is sometimes called the {\it barotropic} EoS or the {\it linear} EoS. However, most dictionaries give the meaning of {\it barotropic} as the property that the pressure depends on the density only, and {\it linear} would include relationships like $p = p_{0} +w\rho$, so we believe that the phrase {\it isothermal} EoS is more appropriate, similar to the isothermal EoS of the classical ideal gas, $pV =$ constant} EoS, (\ref{eos}) is relevant. For example, $w=0$ describes the matter-dominated (or ``pressureless dust") case, since galaxies are taken to behave like the atoms of a cold gas filling the universe, \mbox{$w=1/3$} the radiation-dominated case (e.g. early universe), \mbox{$w<-1/3$} dark energy and \mbox{$w<-1$} phantom energy; concepts that arose recently \cite{de,phantom}, in the context of the acceleration of the expansion of the universe \cite{acceleration-hiZsst,acceleration-SCP}.

  The line element for a static spherically symmetric  spacetime can be written~\cite[Sect.23.2]{mtw} in Schwarzschild (or curvature) coordinates as
\begin{equation}
ds^{2} = -B(r) dt^{2} + A(r) dr^{2} + r^{2} d\Omega^{2}   \label{Schw-ansatz}
\end{equation}
where $d\Omega^{2}$ is the line element for the unit sphere. If the spacetime is sourced by a {\it static} isotropic perfect fluid (SSSPF solutions), one has to also use the staticity of the fluid,  
\begin{equation}
u^{\mu} = u^{0} \delta_{0}^{\mu}  \Longrightarrow u^{0} = B(r)^{-1/2}. \label{staticity}
\end{equation}

However, for most EoS it is very hard to find an analytical solution for the SSSPF problem, that is, functions $B(r)$ and $A(r)$ so that EFE (\ref{ee}) are satisfied [We would like to also warn that such a solution with horizon(s) does {\em not} describe a blackhole (see \cite{revisited,Semiz_on_Kiselev}) since the staticity condition (\ref{staticity}), $u^{\mu} = u^{0} \delta_{0}^{\mu}$ is not valid  inside the horizon].  As stated above, the general solution for the EoS $p=w\rho$ with constant $w$, that is, the spacetime around a spherically symmetric object embedded in dark energy or quintessence\footnote{Some authors use {\it quintessence} as a synonym for  dark energy, while others limit the use of that expression to the range $-1/3>w>-1$.} is not known, even though the EoS looks very simple. Please note that the Kiselev spacetime~\cite{Kiselev} is {\em not} supported by quintessence~\cite{Visser_on_Kiselev,Lake_on_Kiselev,Semiz_on_Kiselev}), since its source is not isotropic.

 Full solutions of the $w=$ constant SSSPF problem are known only for some particular values of $w$: The cases of $w=-1$ and $w=0$ are trivial, the full solution for case $w=-1/3$ was reported in 2001/2002~\cite{CSB}, and the full solution for $w=-1/5$ in 2020/2022~\cite{semiz_w_1/5} (by the present author). This set of $w$ values is consistent with the set conjectured in~\cite{Ivanov} as the set of possible anaytic solutions. The author of~\cite{Ivanov} comes to this conclusion by converting the problem into the Abel equation of second kind, whose integrable cases are known.  The author also points out that the so-called (involutory) Buchdahl transformation~\cite{Buchdahl} connects each $w$ solution to another, in particular, $w=-1$ to $w=-1/5$; hence that author comes close to solving the $w=-1/5$ case (For more detail, and some other aspects of the $w=$ constant problem, see \cite{semiz_w_1/5}).

  On the other hand, the paper~\cite{Liouvillian} claims that $w=-1/6$ should also be integrable, but does not give any solution. The authors base their assertion on the problem being convertible into a Lotka-Volterra differential system whose Liouvillian integrability can be characterized via Darboux theory. In the present work, we find that their conclusion is right, and display the explicit $w=-1/6$ solution. It can be reached by the Buchdahl transform of a known solution, as will be shown below.

The Buchdahl transformation \cite{Buchdahl} (see also Appendix of \cite{semiz_w_1/5}) of a general static line element is
\begin{equation}
-f(x^{i}) dt^{2} + g_{ij}(x^{k}) dx^{i}dx^{j} \longrightarrow -\frac{1}{f(x^{i})} dt^{2} + f^{2}(x^{l}) g_{ij}(x^{k}) dx^{i}dx^{j}   \label{Buchdahl_transform}
\end{equation}
where the indices $i,j,k,l$ run from 1 to 3, namely over the space coordinates. Under the same transformation, the SEM tensor transforms as
\begin{equation}
T^{k}_{\;\;j} \longrightarrow \frac{1}{f^{2}(x^{i})} T^{k}_{\;\;j}, \;\;\;\;\;\;
T^{0}_{\;\;0} \longrightarrow -\frac{1}{f^{2}(x^{i})} (T^{0}_{\;\;0} - 2 T^{j}_{\;\;j})   
\label{SEM-Buchdahl-transform}
\end{equation}
If $T^{\mu\nu}$ represents an isotropic perfect fluid, the transformation of the fluid variables $\rho$ and $p$ in (\ref{pfemt}) becomes
\begin{equation}
\rho \longrightarrow -\frac{\rho+6p}{f^{2}(x^{i})}, \;\;\;\;\;\;
p \longrightarrow \frac{p}{f^{2}(x^{i})}.  
\label{rho-p-Buchdahl-transform}
\end{equation}
Let us also note at this point that the transformation (\ref{Buchdahl_transform}) disturbs the form (\ref{Schw-ansatz}), i.e. the new metric is formulated in some different type of coordinate system; see \cite{Semiz_on_Kiselev} for the different {\em coordinate conditions} in the SSS context.

%
%

\section{Solution for w = -1/6}

From eq.(\ref{rho-p-Buchdahl-transform}), it can be seen that $\rho+6p=0$ transforms into $\rho=0$. But the Buchdahl transform is involutory, hence $\rho=0$ will also transform into the $\rho+6p=0$, i.e. $w=-1/6$ solution. The $\rho=0$ solution, interesting in its own right since it demonstrates that pure pressure by itself can gravitate, was first found by Kuchowicz~\cite{kuch} as far as we know, and is given by the line element
\begin{equation}
ds^2 = - f^{2}(r) dt^2 + \frac{dr^2}{1-C/r} + r^2 d\Omega^2.
\label{Kuch68I}
\end{equation}
where 
\begin{equation}
f(r) = r_1^{-2} \left[(2 r^{2}+5Cr-15 C^{2}) + 15 C^{2} \sqrt{1-\frac{C}{r}} \,\ln\left(\sqrt{\frac{r-C}{C_3}}+\sqrt{\frac{r}{C_3}}\right)\right]
\label{Kuch68I_f}
\end{equation}
and for nonzero $C$, we have absorbed the constant $C_1$ of~\cite{revisited} (equivalently $A$ of~\cite{kuch}) into $C_3$, and also introduced $r_1$ to keep $f(r)$ dimensionless for now (see the relevant footnote in~\cite{revisited} for correspondence of the two forms). For vanishing $C$ however, $C_1$ stays, and $f(r)$ becomes $(r^2 + C_1)/r_2^{2}$.

Applying the Buchdahl transform, we obtain
\begin{equation}
ds^2 = - \frac{dt^2}{f^{2}(r)}  + \frac{f^{4}(r)}{1-C/r}dr^2 + f^{4}(r)r^2 d\Omega^2
\label{eq:w=-1/6}
\end{equation}
as the desired $w=-1/6$ SSSPF solution, with SEM tensor functions
\begin{equation}
\rho = - \frac{48}{\kappa \, r_1^2 f^{5}(r)} \;\;\;\;\; {\rm and} \;\;\;\;\; p = \frac{8}{\kappa \, r_1^2 f^{5}(r)}.
\label{-1/6_rho_and_p}
\end{equation}

%
%

\section{Discussion of the spacetimes}\label{Discussion}

The scalar curvature diverges at the roots of $f(r)$, so these $r$ values are singularities. Note that the energy density $\rho$ and pressure $p$ also diverge at these points. These singularities {\em are} points, since the areas of constant-$r$ surfaces vanish at these values.

For correct signature and reality of the square-roots, $(r-C)$, $r$ and $C_3$ must have same sign; therefore the $r$-range between $r=0$ and $r=C$ is excluded. However, this excluded range will span either positive or negative $r$ values depending on the sign of $C$, hence the cases $C>0$, $C=0$ and $C<0$ have to be discussed separately. There are no horizons since $g_{00}$ does not change sign.  The different cases, and their subcases where applicable, are discussed below and displayed in Table \ref{tab}.  

\begin{table}
\caption{Spacetimes in this work. $A$: area, NS: pointlike naked singularity, wh: infinitesimal wormhole. The inner and outer boundaries are given according to te magnitude of the $r$ coordinate, however for the SR(1) and SR(2) spacetimes, the labels should really be switched, since the "outer" boundary is a point whereas the "inner" boundary has nonzero area.} 
\centering
\begin{tabular}{|p{0.08\linewidth}|p{0.06\linewidth}|p{0.13\linewidth}|p{0.16\linewidth}|p{0.13\linewidth}|p{0.15\linewidth}|p{0.12\linewidth}|}
\hline
\multicolumn{3}{|c|}{\textbf{Criteria}} & \textbf{Inner boundary} & \textbf{Outer boundary} & \textbf{Comment} & \textbf{Suggested name} \\ \hline

\multirow{4}{*}{\parbox{\linewidth}{$C=0$, $r$ positive}} & \multicolumn{2}{c|}{$C_1=0$} & $r=0$ (NS)  & \centering $+\infty$ & infinite, not novel  (- also) &  S$\infty$(1)  \\  \cmidrule{2-7}

 & \multicolumn{2}{c|}{$C_1$ positive} & $r=0$ (regular pt) & \centering $+\infty$ & infinite \linebreak (- also) \linebreak (wh also) & 0$\infty$ \linebreak or \linebreak $\infty$0$\infty$ \\ \cmidrule{2-7} 

 & \multirow{2}{*}{\parbox{\linewidth}{$C_1$ negative}} & $r < \sqrt{-C_1}$  & $r=0$ (regular pt) & $r = \sqrt{-C_1}$ (NS)  & compact \linebreak (- also) \linebreak (wh also) & 0S \linebreak or \linebreak S0S   \\ \cmidrule{3-7} 

 & & $r > \sqrt{-C_1}$ & $r=\sqrt{-C_1}$, NS & \centering $+\infty$ & infinite \linebreak (- also) & S$\infty$(2) \\  \hline 

 \multirow{6}{*}{\parbox{\linewidth}{$C$ positive}} & \multirow{2}{*}{$r>C$} & $C<r<r_*$ & $r=C$ (regular surface, nonzero $A$) & $r=r_*$ (NS)  & compact, one boundary unusual & SR(1)   \\ \cmidrule{3-7} 
 
 & & $r>r_*>C$ & $r=r_*$ (NS) & \centering $+\infty$ & infinite & S$\infty$(3)   \\ \cmidrule{2-7} 

 & \multirow{4}{*}{$r<0$ } & $|\alpha|<1$ & $r=0$ (regular surface, nonzero $A$) & \centering $-\infty$ & infinite, one boundary unusual & R$\infty$   \\ \cmidrule{3-7} 

 & & $|\alpha| = 1$  & $ r=0$ (NS)  & \centering $-\infty$ & infinite & S$\infty$(4)   \\ \cmidrule{3-7}

  & & $|\alpha| > 1$ (iii) & $r=0$ (regular surface, nonzero $A$)  & $r=r_*$ (NS)  & compact, one boundary unusual & SR(2)   \\ \cmidrule{3-7}

 & & $|\alpha| > 1$ (iv)  & $r=r_*$ (NS)  & \centering $-\infty$ & infinite & S$\infty$(5)   \\  \hline 
 
$C$ negative & \multicolumn{6}{c|}{\parbox{0.85\linewidth}{The transformation $r \longrightarrow -r$ will map all the spacetimes of this case to the "$C$ positive" spacetimes, and vice versa.}}   \\ \hline 
 
\end{tabular}
\label{tab}
\end{table}

\subsection{The special case of vanishing C}\label{vanishingC}

In this case the line element simplifies greatly, becoming
\begin{equation}
ds^2 = - \left(\frac{r_1^2}{r^2+C_1}\right)^2 dt^2 + \left(\frac{r^2+C_1}{r_1^2}\right)^4 dr^2 + \left(\frac{r^2+C_1}{r_1^2}\right)^4 r^2 d\Omega^2.
\label{C_vanish}
\end{equation}
This line element has a singularity for negative $C_1$ and can be further simplified if $C_1$ vanishes, so we have to distinguish these three cases.

\subsubsection{The extra-special subcase of vanishing $C_1$} \label{s_inf1}

If $C_1$ also vanishes, the line element becomes
\begin{equation}
ds^2 = \frac{1}{r_1^8} \left(-\frac{dt^2}{r^4} + r^8 dr^2 + r^{10} d\Omega^2 \right)
\label{C_C1_vanish}
\end{equation}
after rescaling the time coordinate. Passing to the Schwarzschild or curvature radial coordinate $\bar{r}=r^5/r_1^4$, we obtain
\begin{equation}
ds^2 =  -\frac{r_2^{4/5}}{\bar{r}^{4/5}}dt^2 + \frac{1}{25} d\bar{r}^2 + \bar{r}^{2} d\Omega^2. 
\label{C_C1_vanish_fin}
\end{equation}
This can be recognized as the $w=-1/6$ case of Solution 7a of~\cite{allpoly}, which itself is a special case of the Tolman V metric~\cite{Tolman}, hence is not novel.
For this metric, the scalar curvature, density, pressure all diverge at the origin, so this is an infinite spacetime with a naked singularity at the origin; we suggest using the name S$\infty$ (cf. the spacetime of the same name in Sect.3.1.1 of~\cite{semiz_w_1/5}) for completeness, despite the metric not being novel. If negative $r$ is allowed, an identical spacetime exists in that range. 

\subsubsection{The subcase of positive $C_1$} \label{inf0inf}

In this case, $f(r)$ never vanishes, so the spacetime is infinite, and everywhere regular; we can name it 0$\infty$. Again, if negative $r$ is allowed, an identical spacetime exists in that range; but this time, the negative-$r$ spacetime can be imagined touching its positive-$r$ counterpart at one point, which then can be called $\infty$0$\infty$ spacetime (cf. the spacetime(s) of the same name in Sect.3.1.1 of~\cite{semiz_w_1/5}). 
 
\subsubsection{The subcase of negative $C_1$}  \label{S0SandSinf}
 
Let us call $C_1=-r_0^2$; then there are pointlike singularities at $r=\pm r_0$, as discussed in the beginning of this section.

The region near $r=0$ is regular, almost Minkowskian; therefore the region $0 \leq r \leq r_0$ is a compact, closed spacetime containing a naked singularity, hence a ``0S spacetime'' as in in the last case of Sect.3.1.3 of~\cite{semiz_w_1/5}.

The region $r > r_0$ is an infinite spacetime with a naked singularity at the center, another S$\infty$ type spacetime.

Once again, if one allows negative $r$, there are identical spacetimes in that range; and the closed one can be imagined to be touching its positive-$r$ counterpart at one point (``S0S'').

\subsection{The case of positive C} \label{positiveC}

In this case, we can have $r>C$ or $r<0$, as discussed in the beginning of this section. For better understanding, one can ``simplify'' the line element by introducing $\bar{r}=r/C$ and then rescaling time to get
\begin{subequations}
\begin{equation}
ds^2 = C^6 \left[ - \frac{dt^2}{\bar{f}^{2}(\bar{r})}  + \frac{\bar{f}^{4}(\bar{r})}{1-1/\bar{r}}d\bar{r}^2 + \bar{f}^{4}(r)\bar{r}^2 d\Omega^2 \right]
\label{le_conf+C}
\end{equation}
with
\begin{equation}
\bar{f}(\bar{r}) = (2 \bar{r}^{2}+5\bar{r}-15) + 15 \sqrt{1-\frac{1}{\bar{r}}} \,\ln\left(\sqrt{\frac{\bar{r}-1}{\alpha}}+\sqrt{\frac{\bar{r}}{\alpha}}\right)
\label{fbar+C}
\end{equation}
\label{w=-1/6_conf+C}
\end{subequations}
%

\subsubsection{The subcase of $r > C$.}  \label{sr}

The function $\bar{f}(\bar{r})$ in the metric~(\ref{w=-1/6_conf+C}) starts at -8 at $\bar{r}=1$, and approaches $2 \bar{r}^{2}$ at large $\bar{r}$ values, i.e. has a root (call it $\bar{r}=\bar{r}_*$) whose value depends on $\alpha$ in a nontrivial way. Since $r = r_* = C \bar{r}_*$ is a singularity, the solution~(\ref{w=-1/6_conf+C}) includes two spacetimes, one for $C<r<r_*$ and one for $r>r_*$. Note that $C_3$, hence $\alpha$ must be positive in this subcase according to the discussion in the beginning of this section. 

The $0<C<r<r_*$ spacetime is compact despite the divergence of $g_{11}$ at $r=C$ or $\bar{r}=1$.
That spherical surface has area radius $64 C^3$, and therefore it is more apropriate to take the singular point $r=r_*$ as the center and $r=C$ as the outer boundary. That boundary is not null, therefore it cannot be a horizon; nor is it infinitely far away. At that surface, or before, the spacetime must be matched to another, satisfying another EoS, if supported by a perfect fluid at all. This spacetime may be called SR (R for regular).

The $r>r_*>C>0$ spacetime extends from the central (but $\bar{r}=\bar{r}_*$, not $\bar{r}=0$ or $r=0$) naked singularity to infinity, yet another S$\infty$ type spacetime.

\subsubsection{The subcase of $r < 0$.}  \label{posCneg_r}

 In this subcase, $C_3$, hence $\alpha$ must be negative. The function $\bar{f}(\bar{r}$) has one root (again, call it $\bar{r}=\bar{r}_*$)  if $|\alpha| \geq 1$, no roots if $|\alpha| < 1$; and $\bar{f}(\bar{r})$ diverges as $|\bar{r}|^{-1/2}$ at $\bar{r}=0$ unless $|\alpha|=1$, in which case $\bar{f}(\bar{r})$ vanishes there. Therefore, this solution contains four possible spacetimes: (i) $|\alpha| < 1$ (ii) $|\alpha|=1$ (iii) $|\alpha| > 1$, $r_* < r < 0$ (iv) $|\alpha| > 1$, $r < r_* < 0$:

\paragraph{(i) $|\alpha| < 1$:}
$r=0$ is a spherical surface with radius $|\frac{15}{2} \ln \alpha|^2 \bar{C}^3$ rather than a point, and $\bar{f}(\bar{r})$ is always positive, so we have an infinite spacetime bounded by a regular sphere inside, which may be called an R$\infty$ spacetime.

\paragraph{(ii) $|\alpha|=1$:}
$r=0$ is singular (since $\bar{f}(0)=0$), hence this spacetime is of type S$\infty$.

\paragraph{(iii) $|\alpha| > 1$, $r_* < r < 0$:}
This spacetime is compact, and $\bar{f}(\bar{r})$ is negative. $\bar{r}=\bar{r}_*$ is a naked singularity, and $r=0$ a spherical surface with radius $|\frac{15}{2} \ln \alpha|^2 \bar{C}^3$, so this is another SR type spacetime.

\paragraph{(iv) $|\alpha| > 1$, $r < r_* < 0$:}
This spacetime is the fifth S$\infty$ type one of this work.

\subsection{The case of negative C}\label{negativeC}

 For negative $C$, on the other hand, we can have $r<C$ or $r>0$. We can similarly simplify the line element (this time $\bar{r}=r/|C|$) to get
\begin{subequations}
\begin{equation}
ds^2 = \bar{C}^6 \left[ - \frac{dt^2}{\bar{f}^{2}(\bar{r})}  + \frac{\bar{f}^{4}(\bar{r})}{1+1/\bar{r}}d\bar{r}^2 + \bar{f}^{4}(r)\bar{r}^2 d\Omega^2 \right]
\label{le_conf-C}
\end{equation}
with
\begin{equation}
\bar{f}(\bar{r}) = (2 \bar{r}^{2}-5\bar{r}-15) + 15 \sqrt{1+\frac{1}{\bar{r}}} \,\ln\left(\sqrt{\frac{\bar{r}+1}{\alpha}}+\sqrt{\frac{\bar{r}}{\alpha}}\right)
\label{fbar-C}
\end{equation}
\label{w=-1/6_conf-C}
\end{subequations}
where the definition of $\alpha$ is slightly different from the positive $C$ case, (\ref{w=-1/6_conf+C}). However, it can be easily seen that the transformation $r \longrightarrow -r$, hence $\bar{r} \longrightarrow -\bar{r}$ leads to spacetimes identical to those in Sect.\ref{positiveC}.

\subsection{The conformal (Carter-Penrose) diagrams}

The $w=-1/6$ general line element, (\ref{eq:w=-1/6}), can be written as
\begin{equation}
ds^{2} = \frac {1}{f^2(r)} \left(-dt^{2} + d\bar{R}^{2}  \right)   \label{GenMetricConf1}
\end{equation}
where the angular coordinates have been suppressed and $\bar{R}$ has been defined via 
\begin{equation}
\bar{R} = \int f^3(r) \sqrt{\frac{r}{r-C}} dr, \label{ConfR}
\end{equation}
and $r$ and $f(r)$ in (\ref{GenMetricConf1}) are now considered to be implicit functions of $\bar{R}$ via (\ref{ConfR}). The form (\ref{GenMetricConf1}) allows  construction of conformal diagrams which elucidate causal relationships.

\begin{figure}[h!]
\caption{Possible Penrose diagrams of the S$\infty$ spacetimes. The inner boundary (singularity) is at $r=0$ for S$\infty$(1) and S$\infty$(4); at $r=r_0=\sqrt{-C_1}$ for S$\infty$(2); and at $r=r_*$ (positive and negative, restectively) for S$\infty$(3) and S$\infty$(5). $r=-\infty$ for the outer boundary applies to S$\infty$(4) and S$\infty$(5).} 
\centering
\begin{picture}(300,520)
\put(0,0){\includegraphics[width=0.6 \columnwidth]{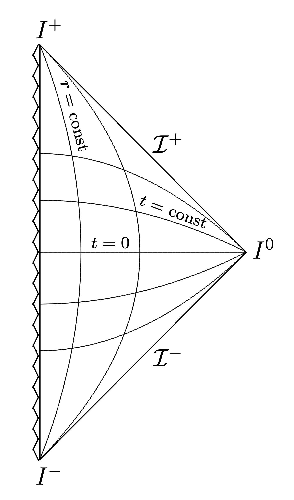}}
\put(20,300){\rotatebox{-90}{$r=0$ or $r_0$ or $r_*$}}
\put(103,418){\rotatebox{-45}{$r=\pm\infty$}}	\put(190,330){\rotatebox{-45}{ $t=\infty$}}
\put(103,85){\rotatebox{45}{$r=\pm\infty$}}	\put(190,170){\rotatebox{45}{$t=-\infty$}}
\end{picture}
\label{fig:s-inf} 
\end{figure}

\paragraph{--} For the S$\infty$(1) spacetime discussed in subsection \ref{s_inf1}, a Penrose diagram can be constructed that is the same as that of the similarly named spacetime in~\cite{semiz_w_1/5}, i.e. Fig.4 of that publication. Here we construct Fig. \ref{fig:s-inf} to show all five S$\infty$ spacetimes together.

\paragraph{--} Similarly, the Penrose diagram of the $\infty$0$\infty$ spacetime of  subsection \ref{inf0inf} can be same as Fig.5 of~\cite{semiz_w_1/5} (or right half of it for  the 0$\infty$ spacetime).

\paragraph{--} The S0S spacetime of  subsection \ref{S0SandSinf} can be represented by the Penrose diagram in Fig.7 of~\cite{semiz_w_1/5} (the 0S spacetime  by half of it); while for the S$\infty$(2) spacetime discussed in that subsection, we can again refer to Fig.\ref{fig:s-inf}.

\paragraph{--} The Penrose diagram of the SR spacetime discussed in subsection \ref{sr}, Fig.\ref{fig:sr}, is somewhat similar to the spacetime called SS in~\cite{semiz_w_1/5}, i.e. Fig.6 of that publication. However, the outer boundary is regular instead of singular, as discussed above. The S$\infty$(3) spacetime of the same subsection is again one of those depicted in Fig.\ref{fig:s-inf}.

\begin{figure}[h!]
\caption{Possible Penrose diagrams for the SR spacetimes. For SR(1), the $r$ values for both boundaries are positive; whereas for SR(2), the the $r$ value for the inner boundary is negative, the outer value is zero (but area does not vanish).} 
\centering
\begin{picture}(300,330)
\thicklines \put(100,25){\line(0,1){270}}  
\multiput(96,26)(0,15){18}{\tiny {\bf \textbackslash}}
\multiput(96,31)(0,15){18}{\tiny {\bf /}}
\put(85,150){\rotatebox{90}{$r=r_*$}}
\qbezier(100,25)(250,160)(100,295)
\put(180,130){\rotatebox{90}{$r=C$ or 0}}
\end{picture}\label{fig:sr} 
\end{figure}

\begin{figure}[h!]
\caption{A Penrose diagram of the R$\infty$ spacetime.} 
\centering
\begin{picture}(300,520)
\put(0,0){\includegraphics[width=0.6 \columnwidth]{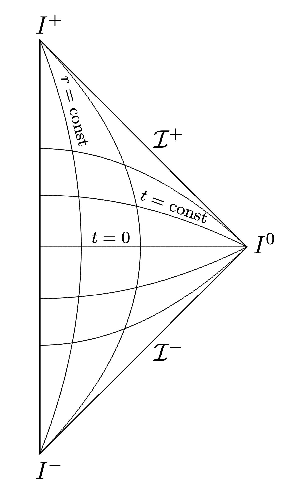}}
\put(22,330){\rotatebox{-90}{$r=0$ (but nonzero area)}}
\put(95,433){\rotatebox{-45}{$r=-\infty$}}	\put(195,333){\rotatebox{-45}{ $t=\infty$}}
\put(97,82){\rotatebox{45}{$r=-\infty$}}	\put(197,182){\rotatebox{45}{$t=-\infty$}}
\end{picture}
\label{fig:r-inf} 
\end{figure}

\paragraph{--} Case (i) of subsection \ref{posCneg_r} gives a Penrose diagram similar to the S$\infty$ cases, except that the inner boundary is not singular; see Fig.~\ref{fig:r-inf}.

\paragraph{--} The cases (ii), (iii) and (iv) of subsection \ref{posCneg_r}, that is, the S$\infty$(4), SR(2) and S$\infty$(5) spacetimes, respectively, are included in Figures \ref{fig:s-inf} and \ref{fig:sr}.

%
%

\section{Summary and final comments}

We have reported the full solution of Einstein's Equations for a perfect fluid source with equation of state $p=-\rho/6$, in the static spherically symmetric case. We discussed all ranges of the parameters of the solution, and the spacetimes these ranges correspond to. This solution verifies and realizes the prediction of~\cite{Liouvillian} and adds to the set of fully solved $w$'s which previously was $\{0,-1,-\frac{1}{3},-\frac{1}{5}\}$ (the first two trivial, the last one by the current author~\cite{semiz_w_1/5}).

The EoS $p = w \rho$ is motivated by studies in stellar structure and in particular, cosmology. In the former, $w$ is usually taken to be positive, in the latter, $w<-\frac{1}{3}$ represents dark energy invoked to ``explain'' the accelerating expansion of the universe. Part of the mystery of the dark energy ``fluid'' is that it violates all (if $w<-1$) or some energy conditions, hence is in some sense considered unphysical or exotic; however, it might not be a real fluid, but a manifestation of a field, for example. Note that, regardless of the value of $w$, the density $\rho$ of dark energy  is implicitly assumed to be positive.

Even though reservations exist about fluids with negative-$w$ or those that violate one or more energy conditions, the fact that they are generally employed in cosmology naturally leads to their consideration as possibilities in other contexts. In particular, nonsingular parts of any SSS solutions are candidates for description of parts --some radial range-- of spherical ``stars''; they would probably have to be matched to some other solutions at the ends of the range. 

Should one still use the energy conditions as guides in evaluating how ``physical'' a spacetime is, a look at eqs. (\ref{-1/6_rho_and_p}) shows that those spacetimes with positive $f(r)$ violate all the energy conditions, in fact, they have negative energy density $\rho$; whereas those with negative $f(r)$ satisfy all of them. So, in one sense, the spacetimes 0S (or S0S), SR(1) and SR(2) are more physical than the other spacetimes in this work. These also happen to be the compact spacetimes, whereas the others are infinite.

The final comments of~\cite{semiz_w_1/5} apply also here: The EoS $p=-\rho/6$ does not have an obvious interpretation, unlike $w=0$, $w=-1$ or $w=-1/3$ (vacuum, $\Lambda$ and  gas of cosmic strings, respectively). Therefore, notwithstanding the mathematical arguments of~\cite{Ivanov} and~\cite{Liouvillian}, it is a bit surprising that the cases $w=-1/5$ and $w=-1/6$ should turn out to be integrable while the general constant $w$ case resists integration. Finally, the diversity of the spacetimes of this case further dampens hopes for the integrability of the mentioned general case.

\end{document}